\documentstyle[multicol,aps,epsf]{revtex}
\begin{document}
\title{Melting of Flux Lines in an Alternating Parallel Current}
\author{Mohammad Kohandel}
\address{Institute for Advanced Studies in Basic Sciences,
Zanjan 45195-159, Iran}
\author{Mehran Kardar}
\address{Department of Physics,
Massachusetts Institute of Technology, Cambridge, Massachusetts, 02139 }
\date{\today}
\maketitle
\begin{abstract}
We use a Langevin equation to examine the  dynamics and fluctuations of a flux line 
(FL) in the presence of an {\it alternating longitudinal current}  $J_{\parallel}(\omega)$.
The magnus and dissipative forces are equated to those resulting from line tension,
confinement in a harmonic cage by neighboring FLs, parallel current, and noise.
The resulting mean-square FL fluctuations are calculated {\it exactly}, and
a Lindemann criterion is then used to obtain a nonequilibrium
`phase diagram' as a function of the magnitude and frequency of $J_{\parallel}(\omega)$.
For zero frequency, the melting temperature of the mixed phase (a lattice, or the
putative ``Bose" or ``Bragg Glass") vanishes at a limiting current.
However, for any finite frequency, there is a non-zero melting temperature.
\vspace{1cm}
\end{abstract}
\begin{multicols}{2}
\section{Introduction}
The phases of flux lines (FLs) in high temperature superconductors are the subject 
of many current experimental and theoretical investigations \cite{Blatter,Brandt1}. 
In the classical (Abrikosov)  picture of type-II superconductors, FLs penetrate a 
clean material for fields $H>H_{c1}$, to form a triangular solid lattice. 
This mean field phase diagram is modified by inclusion of thermal fluctuations, 
or in the presence of various forms of disorder  (such as point or columnar defects), 
which play an important role in high-$T_c$ materials\cite{Larkin,Fisher}.
The essentially classical problem of understanding the equilibrium and
non-equilibrium properties of the resulting vortex matter composed of meandering
FLs, poses a myriad of interesting challenges \cite{Blatter,Brandt1,Nelson1,Kardar}.

There is now growing evidence that the mixed phase of high-$T_c$ materials is a 
``Bose Glass" (BoG)\cite{Nelson1,Blatter}, in which the FLs are pinned by such 
extended (or correlated) imperfections as grain boundaries or columnar pins. 
Since the FLs are pinned to these large defects, 
the BoG has an infinite tilt modulus $c_{44}$.
There is also a cusp-like singularity in its melting temperature  in a field $H_{\perp}$,
perpendicular to the extended defects \cite{Nelson1}.
On the other hand, if the defects are at uncorrelated points 
(such as oxygen impurities)\cite{GL}, there may exist a  
``Bragg Glass" (BrG) phase at low magnetic fields and temperatures \cite{EN}. 
The dislocation free  ``Bragg Glass" has a kind of topological order.
Experiments on BSCCO \cite{Khay} have measured the effects
of both point and columnar imperfections, showing that a melting transition 
exists in the presence of either type of disorder. 
For technical applications, it may be useful to control the melting temperature of 
the mixed phase (BoG or BrG) by application of external parameters. 
One possibility is an alternating longitudinal current $J_{\parallel}(\omega)$,
as in some recent experiments \cite{Danna}.

The response of FLs to $J_{\parallel}$ was studied extensively even prior
to the high-$T_c$ era \cite{Brandt1}.
Since the Lorentz force $\vec F=\phi_0 \vec J \times \vec B$ is absent for
$\vec J \parallel \vec B$, the straight FL is stable. Any fluctuation of the
FL, however, results in a force that causes helical instabilities
at long wavelengths \cite{Brandt1}.
A lattice of FLs is similarly unstable to an infinitesimal $J_{\parallel}$, if
it penetrates the whole sample. As shown by Brandt
\cite{Brandt2}, for a finite penetration depth of current,
a finite $J_{\parallel} \sim {\varepsilon}^{{1}/{4}}$, is
needed to cause instability, where $\varepsilon$ is related to the shear modulus $c_{66}$. The many studies of this subject point to
the importance of surface effects \cite{Brandt1,Brandt2}, and the bewildering
behavior of the unstable FLs.

Brandt has suggested\cite{Brandt1}  that FLs can also be stabilized for infinitesimal
$J_{\parallel}$ due to pinning by disorder. However, this study predates the
introduction of the BoG and BrG phases, which are more ordered than a flux liquid.
We may thus enquire if these phases are stable with respect to $J_{\parallel}$.
More generally, what is the behavior of the flux matter in response to 
an {\it alternating} current $J_{\parallel}(\omega)$?
In what follows, we shall assume a current $J_{\parallel}(\omega)$ that
uniformly penetrates the whole sample, and calculate a `melting temperature' ($T_m$)
based on this assumption. Self-consistency of the treatment
thus requires penetration of $J_{\parallel}$ through the whole sample.
Whether or not the longitudinal penetration depth $\lambda_{\parallel}(
\omega)$ is finite at $T_m$ appears to be unsettled. This will be discussed
in more detail in the next sections. If it does happen that
$\lambda_{\parallel}$ is finite at the transition, our result becomes
applicable only to thin samples with thickness of a few
$\lambda_{\parallel}$, or alternatively, only close to the surface of the
sample. Nonetheless, such a melting at the surface  will undoubtedly 
effect the transport properties of the superconductor, and may
well trigger melting in the bulk of the material.

We use the {\it cage model} to examine the fluctuations of one
representative FL in the confining potential provided by its neighbors.
The collective interactions of many FLs is thus approximated by a 
simple harmonic potential. 
The cage model provides a good description for the BoG
phase in the limit of $B>{\phi_0}/{\lambda_{\perp}^2}$ 
as argued in Ref.~\cite{Nelson1}, and is also applicable to the BrG, 
as discussed by Ertas and Nelson \cite{EN}.
Basically, there is evidence of some reasonably long, but possibly finite range,
translational order in the flux phase of high temperature materials, despite
the presence of disorder \cite{Nelson1,GL}. 
The cage model is a valuable tool for estimating the limits of such order,
and how it is destroyed by fluctuations. 
More recently, the cage model was used by Goldschmidt \cite{Gold},
in conjunction with a replica variational method, to compare and contrast the effect
of point and columnar disorders on the melting temperature.

So far the cage model has only been used\cite{Nelson1,EN} to study melting 
due to thermal fluctuations or impurities under {\it equilibrium} conditions. 
The Lindemann criterion, however, has been applied to study
the {\it non-equilibrium} melting dynamics of the vortex lattice moving in
a random medium due to a transverse current\cite{Vinokur}.
In the same spirit, we use the Lindemann criterion, in conjunction with the
cage model, to examine the destruction of order  induced by an
alternating parallel current. Inclusion of a time dependent current  necessarily
implies the need to go beyond the standard treatments of a Hamiltonian\cite{Nelson1,Fingel}.
Instead, we work directly with a Langevin equation for the motion of a FL, enabling
us to estimate fluctuations in the presence of $J_{\parallel}(\omega)$.

The rest of this paper is organized as follows.
In Sec.~II, we introduce the cage model and construct an equation of motion for the FL
in the presence of $J_{\parallel}(\omega)$. Using a Lindemann criterion, we then estimate
a melting temperature $T_m$, and examine its dependence on the amplitude
and frequency of $J_{\parallel}$. Application of this result to BoG and BrG are discussed
in Secs.~III and IV respectively. The conclusions appear in sec.~V, and
details of the steady state distribution for $\omega=0$ are placed in the Appendix.

\section{The Cage Model}
In the spirit of the Einstein model for vibrations of a solid, the ``cage model"
approximates the system of many interacting FLs by that of a single FL confined
by its neighbors to a harmonic potential $k r^2/2$. The spring constant $k$, is
approximated in the dense limit of $B>H_{c_1}$, by $k \approx {\varepsilon_0}/{a^2}$;
where $\varepsilon_0=(\phi_0/{4 \pi \lambda})^2$, and $\phi_0=hc/2e$.
The spacing of flux lines is indicated by $a$, while
$\lambda$ is the penetration depth \cite{Blatter}. The energy of the FL is then given by
the Hamiltonian
\begin{equation}\label{Ham}
{\cal H}=\int dz \left[\frac{\varepsilon_1}{2} \left(\frac{\partial \vec r}{\partial z}\right)^2+
\frac{k}{2}\,\vec r\,^2\right],
\end{equation}
where $\vec r\,(z)$ is a two-component vector describing  the FL configuration 
($\hat z\parallel \vec B\,$). In the addition to the harmonic potential, there is an effective
line tension $\varepsilon_1=\varepsilon_0/\gamma^2$, 
where $\gamma^2=m_z/m_{\perp} \gg 1$ is the mass anisotropy.

In the Langevin approach, the dynamics of the field $\vec r\,(z,t)$ is governed by
the equation of motion
\begin{equation}\label{eomA}
\dot {\vec r}=- \mu \frac{\delta {\cal H}}{\delta \vec r} +\vec \eta(z,t)=
\mu \left(\varepsilon_1 {\partial_z}^2 \vec r-k \vec r\,\right)+\vec {\eta}(z,t),
\end{equation}
where $\mu$ is the mobility of the FL, estimated by the Bardeen-Stephen \cite{Bardeen}
expression $\mu={\rho_n}/(\phi_0 H_{c_2})$, where $\rho_n$ the normal state
resistivity of the material. 
The random noise $\vec\eta$ has zero mean, and is uncorrelated at different
points and times, such that
\begin{equation}\label{noiseA}
\left\langle\eta_\alpha(z,t)\eta_\beta(z^\prime,t^\prime)\right\rangle=
2D\delta_{\alpha\beta}\delta(z-z^\prime)\delta(t-t^\prime).
\end{equation}
To ensure that in the steady state the fluctuations are governed by the Boltzmann
weight of the Hamiltonian ${\cal H}$ at a temperature $T$, we must satisfy the relation 
\begin{equation}\label{noiseB}
D=\mu k_BT.
\end{equation}

The actual  dynamics of a FL can be more complicated.
Even if restricted to local terms, the symmetries of the system allow the inclusion
of a {\it Magnus force}, generalizing Eq.~(\ref{eomA})  to
\begin{equation}\label{eomB}
\dot{\vec  r}+\alpha \hat z \times \dot{\vec  r}=
- \mu\frac{\delta H}{\delta \vec r} +\vec \eta(z,t),
\end{equation}
where $\mu$ is again the  mobility. The parameter $\alpha$ is related to 
the Hall angle in the normal phase by $\alpha=\tan \theta_H$  \cite{Blatter},
and expresses the tendency of the FL to drift perpendicular to the applied force.
In the Appendix, we prove that even for this more general equation, the requirement
of thermal equilibrium is still satisfied if Eq.~(\ref{noiseB}) holds.

A time dependent current $\vec J(t)$, destroys thermal equilibrium and can not
be placed in the Hamiltonian. Such a current generates a Lorentz force
$\vec F=\phi_0  \vec J\times\vec t$, where $\vec t$ is the unit tangent vector.
In the case of a current $\vec J \parallel \hat z$, the (linearized) force is 
$\phi_0 J_z(t) \hat z \times \partial_z \vec r$. 
We shall assume that close to equilibrium this force can be simply added to the
time evolution in Eq.~(\ref{eomB}). Thus we shall study the equation of motion
\begin{equation}\label{eomR}
\dot{\vec  r}+\alpha \hat z \times \dot{\vec r}=\mu
\left[\varepsilon_1 {\partial_z}^2 \vec r-k \vec r+\phi_0 J_z(t)
\hat z \times \partial_z \vec r\,\right]+\vec {\eta},
\end{equation}
while keeping the fluctuations of the noise related to ambient 
temperature by Eqs.~(\ref{noiseA}) and (\ref{noiseB}).

By decomposing the vector $\vec r(z,t)$ into its two components $r_1$ and $r_2$,
and using the transformation $R(z,t)=r_1(z,t)+i r_2(z,t)$, 
Eq.~(\ref{eomR}) can be written as
\begin{equation}\label{eomC}
\left(1+i \alpha\right)\dot R=\mu\left[\varepsilon_1 {\partial_z}^2 R-k R+i \phi_0
J_z(t) \partial_z R\right]+\tilde {\eta} ,
\end{equation}
where $\tilde \eta=\eta_1+i \eta_2$. After transforming to Fourier
space, we obtain
\begin{equation}\label{eomF}
\left(1+i \alpha\right)\dot R+\mu\left[\varepsilon_1 q^2+k+\phi_0 J_z(t)q\right]R=
\tilde \eta(q,t),
\end{equation}
whose solution is given by
\begin{eqnarray}\label{Rqt}
R(q,t)&=&\int_{0}^{t} dt^{\prime}\,\frac{\tilde \eta(q,t^{\prime})}{1+i \alpha}
\times   \\
&\exp&\left\{-\frac{\mu}{1+i \alpha} \int_{t^{\prime}}^{t} d\tau
\left[\varepsilon_1 q^2+k+\phi_0 J_z(\tau)q\right]\right\},  \nonumber
\end{eqnarray}
where we have assumed that $R(q,0)=0$. 

From Eq.~(\ref{noiseA}), we obtain that the Fourier components of noise satisfy
\begin{equation}\label{noiseF}
\left\langle\tilde \eta (q,t) \tilde \eta ^*(q^{\prime},t^{\prime})\right\rangle=
4 D (2 \pi)  \delta(q-q^{\prime})\delta(t-t^{\prime}).
\end{equation}
Using  Eqs. (\ref{Rqt}) and (\ref{noiseF})  we find,
\begin{eqnarray}\label{Rqt1}
\left\langle{\left| R(q,t)\right|}\,^2\right\rangle&=&\int_{0}^{t} dt^{\prime}\frac{4 DL_z}{1+{\alpha}^2}
\times  \\
&\exp&\left\{\frac{-2 \mu}{1+{\alpha}^2} \int_{t^{\prime}}^{t} d\tau
\left[\varepsilon_1 q^2+k+\phi_0 J_z(\tau)q\right]\right\},  \nonumber 
\end{eqnarray}
where we have set $2\pi \delta(0)=L_z$, for a sample of length $L_z$ 
in the $z$-direction. 

We shall henceforth consider an alternating current, $J_z=j_z \cos(\omega t)$,
for which the integral in the exponent can be easily performed to yield
\begin{eqnarray}\label{Rqt2}
&&\left\langle{\left| R(q,t)\right|}\,^2\right\rangle=\int_{0}^{t} d t^{\prime}
\frac{4DL_z}{1+{\alpha}^2}\exp\Bigg\{\frac{-2 \mu}{1+{\alpha}^2}
\times   \\
&&\left [\left(\varepsilon_1 q^2+k\right)
(t-t')+\phi_0 j_zq\frac{\sin(\omega t)-\sin(\omega t')}{\omega}\right]\Bigg\}.   \nonumber
\end{eqnarray}
Note that the mean-square fluctuations have an oscillatory dependence on
the time $t$, even in the long-time limit. However, to calculate the fluctuation-induced
melting temperature we have to consider the behavior over the entire cycle.
In particular, we shall focus on the {\it maximal fluctuations} which occur when
$\omega t$ is a multiple of $\pi$\cite{max}. 
After setting $t^{\prime}=t-\tau'$, we obtain in the long-time limit
\begin{eqnarray}\label{Rqt3}
\left\langle{\left| R_{\rm max}(q)\right|}\,^2\right\rangle&=&\frac{4 D L_z}
{1+{\alpha}^2} \int_{0}^{\infty} d \tau'
\times   \\
&\exp&\left\{\frac{-2 \mu  \tau'}{1+{\alpha}^2}\left[\varepsilon_1 q^2+k\pm
\frac{\phi_0 j_z q}{\omega  \tau'} \sin(\omega  \tau')\right]\right\}.   \nonumber
\end{eqnarray}

The extent of FL fluctuations in position-space is given by
\begin{equation}
I\equiv\int_{0}^{L_z} {dz\over L_z} \left\langle r_1(z)^2+r_2(z)^2\right\rangle =
\frac{1}{L_z} \int \frac{dq}{2 \pi} \left\langle\left| R(q)\right|^2\right\rangle .
\end{equation}
Following simple changes of variables, the maximal extent of fluctuations is
obtained from Eq.~(\ref{Rqt3}) as
\begin{equation}\label{MSA}
I=I_c \int_{0}^{\infty} \frac{dx}{\sqrt {x}}
\exp\left\{-x\left [1-\left(\frac{j_z}{j_c}\right)^2 \left(\frac{\sin({\omega x}/{\omega_c})}
{{\omega x}/{\omega_c}}\right)^2\right]\right\}, 
\end{equation}
where $I_c=D/\mu\sqrt { \pi k \varepsilon_1} \approx
D\phi_0 H_{c_2} a \gamma /( \sqrt{\pi} \rho_n \varepsilon_0)$,
$j_c=\sqrt{{4 k \varepsilon_1}/{{\phi_0}^2}}
\approx 2 \varepsilon_0 /(a \gamma \phi_0)$, and $\omega_c={2 \mu k}/
(1+\alpha^2)={2 \rho_n \varepsilon_0}/{(a^2 \phi_0 H_{c_2}(1+\alpha^2))}$.
Using Eq. (\ref{noiseB}), we find that  $I_c$ depends on the temperature as 
$I_c=k_BT/ \sqrt{ \pi k \varepsilon_1}$, but is independent of $\alpha$,
while $\omega_c$ does depend on $\alpha$.

According to the Lindemann criterion, melting occurs when 
$\sqrt{\left\langle \vec r\,^2\right\rangle} \approx c_L a$,
where $c_L$ is a constant of the order of unity. Using this criterion, we can find the
dependence of $T_m$ on $j_z $ and $\omega$. For $\omega=0$ 
the integral in Eq.~(\ref{MSA}) is given by
$I=I_c\sqrt{{\pi}/[1-(j_z/j_c)^2]}$ for $j_z<j_c$, and
diverges for $j_z>j_c$. Therefore, the melting temperature is
\begin{equation}\label{MT}
T_m=T_m^0\left[1-\left(\frac{j_z}{j_c}\right)^2\right]^{{1}/{2}} , \qquad  {\rm for}
\qquad  j_z<j_c,
\end{equation}
where $k_BT_m^0=a^2 {c_L}^2  \sqrt{ k \varepsilon_1}=
a c_L^2 \varepsilon_0 /\gamma$.
Taking, as parameters for BSCCO, $\gamma=50$, $H_{c_2}=100 T$,
$\varepsilon_0=100(K/ \AA)$, $a=225\AA (B\approx4T)$, $\rho_n=
8 \times {10}^{-3} (\Omega-cm)$, and $\alpha=1.75$, leads
to $j_c \approx 40 (A/m^2)$, and $\omega_c \approx 17 kHz$.
Finally, the choice of $c_L=0.2$, leads to $T_m^0 \approx 18^o K$, which is the
estimate for the melting temperature in the absence of  a current $J_z$. 
(Similarly, the parameters $\gamma\approx 7$ and $\varepsilon_0\approx 50$
appropriate to YBCO, yield $T_m^0 \approx 64^o K$.)

The divergence of the integral (for $j_z>j_c$) is removed at any finite
frequency $\omega$, leading to a non-zero melting temperature. The
behavior of $T_m$ for different values of $j_z$ and $\omega$ (obtained
from numerical integration) is shown in Fig.~1.
Due to the onset of helical instabilities for $\omega=0$, the melting temperature
vanishes for $j>j_c$. 
However, at any finite frequency the system retains its stability for all
currents, although fluctuations are enhanced.  
The cage model provides an estimate for the destruction of translational
order by these fluctuations.
\begin{figure}\label{Fig}
\epsfxsize=9truecm 
\centerline{\epsfbox{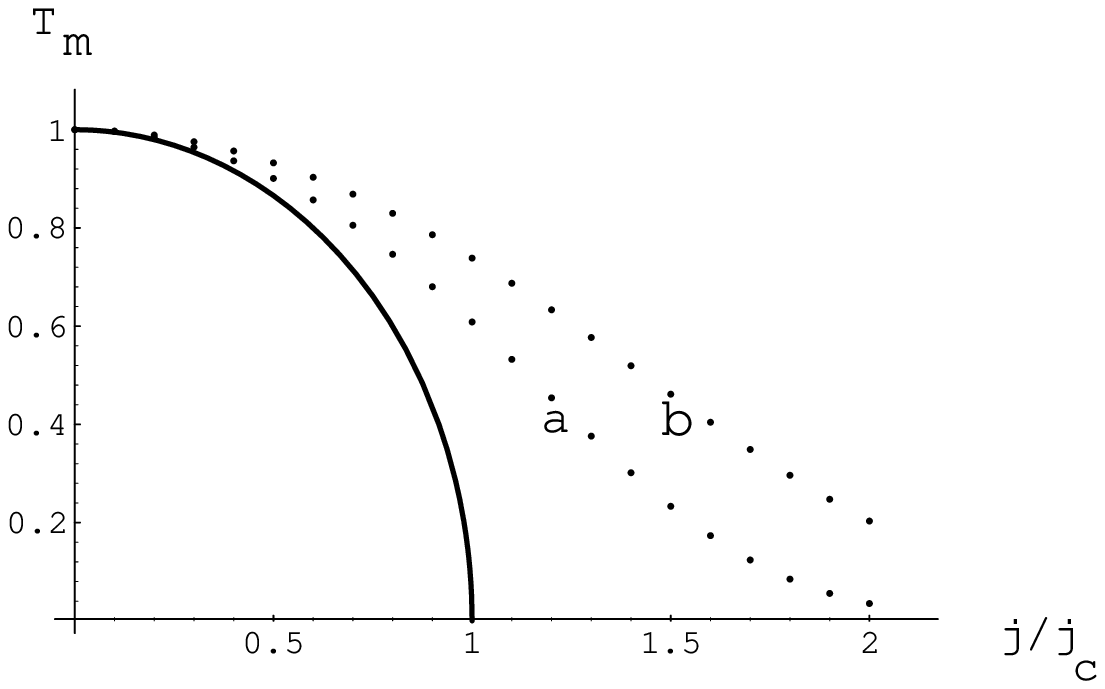}} 
\end{figure}
{\small Fig. 1 Dependence of the melting temperature on $j_z$ and $\omega$.
The solid line is for $\omega=0$, (a) for ${\omega}/{\omega_c}=0.5$,
and (b) for ${\omega}/{\omega_c}=1$.}

\section{The Bose Glass}
To what extent is the previous discussion of a single FL in a cage applicable
to the disordering of the flux phases of high temperature superconductors?
The susceptibility of FLs to pinning by the many sources of impurities
in these materials destroys the long-range translational order of a FL 
lattice\cite{Larkin}, replacing it by one of a number of potential glassy phases.
These glass phases have vanishing linear resistivity because of the
presence of diverging energy barriers at low current density\cite{Fisher}.
From the theoretical perspective, the best understood type of glass phase results 
from correlated disorder, such as parallel columnar pins or grain boundaries. 
Because of a mapping\cite{Nelson1} from this anisotropic system to superfluid 
bosons on a disordered substrate, the resulting phase is called a Bose Glass (BoG).
Due to its anisotropy, the BoG phase has a distinct signature in its conductivity
as a function of orientation.
While the majority of transport experiments are carried out with currents
perpendicular to the magnetic field, recent experiments\cite{Danna} 
measuring electrical transport parallel to the field direction, appear to lend
support to the BoG theory.

As shown by Nelson and Vinokur\cite{Nelson1}, the BoG phase is distinct from
a liquid of FLs, and characterized by a reasonably long translational order 
despite the presence of columnar defects. 
The destruction of this order upon increasing temperature (or magnetic field)
is reminiscent of melting, occuring at  an irreversibility line which is typically at 
higher temperatures than that of pure lattice melting.
It has been argued in Refs.\cite{Nelson1,Radz} that the cage model provides 
a good description for this transition when the magnetic 
field is greater than $B_{\phi} ={\phi_0}/{\lambda_{\perp}^2}$,
which is about $1 T$ for BSCCO.
The results of Fig.~1 should thus apply to the BoG phase in this region, 
establishing the limits of its stability in the presence of a longitudinal current. 

For $j_z=0$, Eq.~(\ref{MT}) reproduces the equilibrium melting temperature of
$T_m=T_m^0= a c_L^2 \varepsilon_0 /\gamma$, as in 
previous studies\cite{Nelson1}.
However, applying the cage model at  finite $J_\parallel(\omega)$ assumes 
that the current uniformly penetrates the whole  sample.
Self-consistency then requires that the current penetration depth should diverge
at the melting temperature; an issue which is not quite clear at present.
Ref.~\cite{Nelson1} claims that when $T \rightarrow T_{BoG}^{-}$,
the penetration length goes to infinity.
However, Ref.~\cite{RN} finds that (for $\omega \rightarrow 0$) the
penetration length diverges as $T \rightarrow T_{BoG}^{-}$
for a perpendicular current,
but remains finite for parallel currents. 
(Note that unlike usual melting, the disordering of
the BoG phase occurs through a continuous phase transition.)
The  AC current also penetrates the sample at the scale of a few $\lambda$
for the relatively (to the gap) low frequencies close to $\omega_c$.
If the current does not fully penetrate the sample, the bulk of the material 
may remain in the BoG phase, while the surface layer melts.
The melting of the FLs at the surface will however profoundly effect the
transport properties of the material, even if no structural change is detected
in the bulk.

\section{The Bragg Glass}
The nature of glassy phases in the case of isotropic disorder (e.g. point defects
due to oxygen vacancies) is more complex.
One proposal is for a ``gauge glass" which assumes complete destruction of
the lattice order.
Another possibility is a dislocation-free Bragg Glass (BrG)
at low temperatures and magnetic fields.
The BrG maintains the topological order of a lattice, and the corresponding
Bragg peaks in scattering experiments\cite{GL}.
Ref.~\cite{EN} argues that due to its similarity to a solid,
interactions of  a FL with its nearest neighbors in the BrG
are well captured by the cage model.
Therefore, one can again apply our results to the BrG phase, 
and the phase diagram of Fig.~1 can also be used to establish the
limits of stability of this phase in the presence of a current $J_\parallel(\omega)$.

In the limit of  $j_z=0$, the equilibrium melting temperature of
$T_m=2 a \varepsilon_0c_L^2/ \gamma$ from Eq.~(\ref{MT})  is  consistent with 
the result of $T_m= 4 a^3 \sqrt{c_{66} c_{44}} c_L^2$ in Ref.\cite{GL} .
The latter is obtained using replica and variational methods,
and coincides with the former if we set  $c_{66}=\varepsilon_0/4 a^2$
and $c_{44}=\varepsilon_0/\gamma^2 a^2$ (single vortex contributions).
The provisos presented before regarding the penetration of the alternating
current into the bulk also apply to this case. However, the melting of the
BrG is thought to be first order, excluding the possibility of diverging
penetration depths at the transition.

\section{Conclusions}
In this paper we examine the response of a single FL in a harmonic
cage to a (destabilizing) parallel alternating current $J_{\parallel} (\omega)$.
A Langevin equation is used to track fluctuations of the FL in
the presence of both $J_{\parallel} (\omega)$ and thermal noise at a temperature $T$.
The harmonic cage is frequently used to mimic the confinement of
a FL by its neighbors in ordered flux phases.
The equilibrium melting temperature (for the destruction of such order)
is then estimated by the Lindemann criterion which equates the extent of
fluctuations to a fraction of the FL spacing.
While the system is no longer at equilibrium in the presence of an alternating current,
we can still ask if the FL order is maintained in the steady state.
In regions where the current penetrates the sample uniformly, we again
apply the Lindemann criterion to answer this question.

Our primary results are summarized by Fig.~1:
(1) At zero temperature and frequency, the FLs are stable up to a current
$j_c$ which is related to the strength of the confining potential.
(2) A uniform current ($\omega=0$) decreases the energy cost of long 
wave-length deformations, resulting in larger FL fluctuations in 
response to thermal noise. The melting temperature $T_m(J_\parallel)$
is thus depressed, eventually going to zero for $J_\parallel=j_c$.
(3) The diverging response of the FL is removed at any finite frequency
$\omega$. Thus while the transition temperature is reduced 
by the alternating current, it remains finite. A substantial reduction
of the transition temperature occurs only at frequencies less than
a characteristic value of $\omega_c$ related to the FL mobility (and
hence to the resistivity in the normal state).
For BSCCO we estimate $j_c \approx 40 (A/m^2)$, and 
$\omega_c \approx 17 kHz$.

While the Lindemann criterion is usually applied to a solid phase, it should
be equally applicable to any of the glassy phases which have a substantial
degree of translational order. 
Our results are thus also relevant to the dislocation free BrG, as well as to
the BoG in the dense limit.
While the pinning of FLs by disorder stabilizes the glassy phases to the current,
we note that the ordered lattice in the absence of impurities is in fact 
unstable for even infinitesimal $J_{\parallel}(\omega=0)$.
It would be interesting to examine the stability of a collection of FLs in
the presence of both impurities and parallel current by extending
previous replica variational calculations\cite{Gold}.
Even the fluctuations of a single FL in the absence of a cage
to the combined effects of current and randomness is an 
interesting topic which can be potentially studies by computer simulations.
Another potential extension is to study the depinning
transition of the FL at zero temperature by methods such as in
Refs.\cite{Kardar,Tang}.

Finally we note that a related situation arises in the context of
chiral polymer crystals \cite{Kamien}.
Long polymers in dense solution often crystallize into a hexagonal
columnar phase. 
When the polymers are chiral, a bias towards
cholesteric twist competes with braiding along an average direction. 
If  the chirality is weak, a defect free hexagonal columnar phase persists, as
in the Meissner phase. 
When the chirality is strong, screw dislocations proliferate, leading to either a 
tilt grain boundary phase or a new state with twisted bond order. 
This new phase may be similar to a flux lattice
in the presence of a parallel current \cite{Kamien}.

We have benefited from discussions with Y. Goldsmith, R. Golestanian, 
M. R. H. Khajehpour, and D.R. Nelson. 
M. Kohandel acknowledges support from the Institute for
Advanced Studies in Basic Sciences, Gava Zang, Zanjan, Iran.
M. Kardar is supported by the NSF through grant number  DMR-93-03667.
\\
\begin{center}
{\bf Appendix}
\end{center}

Despite the differences in  Eq.~(\ref{eomR}) from a standard Langevin
equation, the calculated averages of $\left\langle r^2 \right\rangle$ for 
$\omega=0$ are  those from the Boltzmann weight of a Hamiltonian
\begin{equation}\label{Heff}
{\cal H}=\int dz\left[\frac{\varepsilon_1}{2}
\left(\partial_z \vec r\,\right)^2 -\frac{k}{2} {\vec r\,}^2    
+\frac{\phi_0 j_z}{2}\left(r_2 \partial_z r_1-r_1 \partial_z r_2\right)\right].
\end{equation}
First, we note that a uniform current can be absorbed into an effective 
Hamiltonian as also found in Ref.\cite{Nelson1}.
Second, it can be shown quite generally that a Langevin equation
of the form
\begin{equation}\label{LangevinG}
\left( \delta_{ij}-\alpha\epsilon_{ij} \right)r_j=
-\mu\frac{\delta {\cal H}}{\delta r_i}+\eta_i,
\end{equation}
 where $\epsilon_{ij}$ is the anti-symmetric tensor, has a stationary
state in the form of a Boltzmann weight $\rho_0\propto\exp\left[-{\cal H}/k_BT\right]$,
provided that the correlations of the noise satisfy the condition
in Eq.~(\ref{noiseA}).
For other examples of fluctuation--dissipation conditions in modified
Langevin equations, see Ref.\cite{Zia}.

\end{multicols}

\begin{references}
\bibitem{max}
Looking at some other measure, such as the average fluctuations over
a cycle, should not change the results qualitatively.
\end{references}
\end{document}